\documentclass[12pt]{article}
\pdfoutput=1
\usepackage{epsf}
\usepackage{amsmath,amssymb}

\usepackage{graphicx}
\usepackage{comment}
\usepackage{mathrsfs}

\begin{document}

\newcommand{\lsim}{\stackrel{<}{_\sim}}
\newcommand{\gsim}{\stackrel{>}{_\sim}}
\newcommand{\mathhyphen}{\mathchar"712D}

\newcommand{\rem}[1]{{$\spadesuit$\bf #1$\spadesuit$}}

\renewcommand{\theequation}{\thesection.\arabic{equation}}

\renewcommand{\thefootnote}{\fnsymbol{footnote}}
\setcounter{footnote}{0}

\begin{titlepage}

\begin{center}

\hfill November, 2020

\vskip .5in

{\Large\bf
  Particle Production from Oscillating Scalar Field  \\[2mm]
  and Consistency of Boltzmann Equation
  }

\vskip .5in

{\large
  Takeo Moroi and Wen Yin
}

\vskip 0.25in
{\em Department of Physics, University of Tokyo, Tokyo 113-0033, Japan}

\end{center}

\vskip .3in

\begin{abstract}

  Boltzmann equation plays important roles in particle cosmology in
  studying the evolution of distribution functions (also called as
  occupation numbers) of various particles.  For the case of the decay
  of a scalar condensation $\phi$ into a pair of scalar particles
  (called $\chi$), we point out that the system may not be well
  described by the Boltzmann equation when the occupation number of
  $\chi$ becomes large even in the so-called narrow resonance regime.
  We study the particle production including the possible enhancement
  due to a large occupation number of the final state particle, known
  as the stimulated emission or the parametric resonance.  Based on
  the quantum field theory (QFT), we derive a set of equations which
  directly govern the evolution of the distribution function of
  $\chi$.  Comparing the results of the QFT calculation and those from
  the Boltzmann equation, we find non-agreements in some cases.  In
  particular, in the expanding Universe, the occupation number of
  $\chi$ based on the QFT may differ by many orders of magnitude from
  that from the Boltzmann equation.  We also discuss a possible
  relation between the evolution equations based on the QFT and the
  Boltzmann equation.

\end{abstract}

\end{titlepage}

\setcounter{page}{1}
\renewcommand{\thefootnote}{\#\arabic{footnote}}
\setcounter{footnote}{0}

\section{Introduction}
\setcounter{equation}{0}

Boltzmann equation is widely used as a tool to study the evolution
of the momentum distribution of multiple states.  The success of the
standard cosmology or $\Lambda$CDM essentially relies on the use of
the Boltzmann equation in an expanding Universe.  In cosmology,
however, the stimulated emission or Pauli blocking effect is usually
omitted (see, for example, \cite{Kolb:1990vq, Mukhanov:2005sc}).

Scalar fields may form condensation; examples of such scalar fields
include inflaton \cite{Starobinsky:1980te, Guth:1980zm, Sato:1980yn,
  Linde:1981mu, Albrecht:1982wi}, curvaton \cite{Enqvist:2001zp,
  Lyth:2001nq, Moroi:2001ct}, axion \cite{Preskill:1982cy,
  Abbott:1982af, Dine:1982ah, Graham:2018jyp, Guth:2018hsa},
Affleck-Dine field for the baryogenesis \cite{Affleck:1984fy}, and so
on.  (Hereafter, such a scalar condensation is called $\phi$.)  If
$\phi$ can decay into a pair of bosonic particles (called $\chi$) as
$\phi \to \chi\chi$, the effects of the stimulated emission can be
important since the daughter particles may be enormously populated.  In
particular, recently, a new mechanism of producing bosonic dark matter
from the inflaton decay has been proposed \cite{Moroi:2020has}, in
which a stimulated emission of the bosonic dark matter plays an
important role.

Such systems have been studied by employing the Boltzmann equation or
in the context of the parametric resonance \cite{Traschen:1990sw,
  Kofman:1994rk, Shtanov:1994ce, Yoshimura:1995gc, Kasuya:1996np,
  Kofman:1997yn, Mukhanov:2005sc, Dufaux:2006ee, Matsumoto:2007rd,
  Asaka:2010kv, Mukaida:2013xxa, Kitajima:2017peg, Amin:2019qrx, Garcia:2018wtq,Agrawal:2018vin, Dror:2018pdh, Co:2018lka,  Kaneta:2019zgw,  Lozanov:2019jxc,
  Alonso-Alvarez:2019ssa}.  In particular, particle production from
an oscillating scalar field has been intensively investigated,
particularly using the Mathieu equation \cite{Mathieu, Mathieu2}. In
the previous studies, the evolution equation of the expectation value
of the field operator (which we call wave function) for the
final-state particle is converted to the Mathieu equation, based on
which the occupation number of the final state particle has been
estimated.  Then, it has been shown that there exist resonance bands
and that the occupation numbers in the resonance bands grow
exponentially.  In the broad resonance region, it has been known that
the particle production is non-perturbative and that the process
cannot be described by the Boltzmann equation.  On the contrary,
sometimes it has been argued that the Boltzmann equation can provide a
proper description in the narrow resonance regime.

In this paper, we study the particle production from an oscillating
scalar field in the narrow resonance regime.  We pay particular
attention to the relation between the results based on the Boltzmann
equation and those from the quantum field theory (QFT).  Even though
the effects of the stimulated emission may be included in the
Boltzmann equation \cite{Kolb:1990vq, Mukhanov:2005sc}, it is unclear
whether the Boltzmann equation can properly describe the particle
production from the oscillating scalar field from the QFT point of
view, in particular, the parametric resonance. Based on the QFT, we
derive a set of equations which directly govern the evolution of the
occupation number of the final state particle $\chi$ with an
oscillating $\phi$ background.  We will see that the results of our
evolution equations are in good agreement with those from the
conventional approach with the Mathieu equation.  Then, we compare the
occupation number obtained from the QFT calculation with that from the
Boltzmann equation.  We will see that the results from the QFT and the
Boltzmann equation may differ even in the narrow resonance regime when
the occupation number becomes larger than $\sim 1$.  In particular, in
the expanding Universe, growth factors derived from two
approaches may differ many orders of magnitude.  We also discuss how
our evolution equations can be related to the conventional Boltzmann
equation and consider a possible explanation of the discrepancy.

This paper is organized as follows.  In Section \ref{sec:setup}, the
setup of our analysis is given.  In Section \ref{sec:flat}, we derive
the evolution equations and discuss particle production in the flat
spacetime.  In Section \ref{sec:cosmo}, we study particle production
taking into account the cosmic expansion.  Section \ref{sec:summary}
is devoted to the summary of this paper.

\section{Setup}
\label{sec:setup}
\setcounter{equation}{0}

Here, we study particle production from an oscillating real scalar
field $\phi$, whose mass is $m_\phi$.  We concentrate on the case that
$\phi$ is spatially homogeneous and the scalar field $\phi$ depends
only on time $t$.  We consider the timescale much shorter than the
dissipation of the motion of $\phi$ and the back reaction.  Then,
assuming that the potential of $\phi$ is well approximated by a
parabolic one, the motion of $\phi$ is described as
\begin{align}
  \phi (t) = \bar{\phi} \cos m_\phi t,
  \label{phioscillation}
\end{align}
where $\bar{\phi}$ is the amplitude.  The number density of $\phi$ is
given by
\begin{align}
  n_\phi = \frac{1}{2} m_\phi \bar{\phi}^2.
\end{align}

The scalar field $\phi$ is assumed to interact with a real scalar
field $\chi$.  Although our discussion holds as far as the interaction
Hamiltonian has the form of Eq.\ \eqref{Hint} given below, we consider
the following interaction Hamiltonian to make our discussion concrete:
\begin{align}
  H_{\rm int}= A \int d^3 x \phi \chi^2,
  \label{Hphichichi}
\end{align}
where $A$ is a dimension 1 coupling constant.  With the above interaction
Hamiltonian, the decay rate for the process of
$\phi\rightarrow\chi\chi$ is obtained as
\begin{align}
  \Gamma_{\phi \rightarrow \chi \chi }^{(0)} =
  \frac{A^2}{8\pi m_\phi} \frac{2p_\chi}{m_\phi},
  \label{Gamma(phi->chichi)}
\end{align}
where $p_\chi$ is the three-momentum of $\chi$
produced by the decay:
\begin{align}
  p_\chi \equiv \frac{1}{2} m_\phi \sqrt{1-\frac{4m_\chi^2}{m_\phi^2}},
\end{align}
with $m_\chi$ being the mass of $\chi$.  In
Eq.\ \eqref{Gamma(phi->chichi)}, the superscript ``(0)'' indicates
that $\Gamma_{\phi \rightarrow \chi \chi }^{(0)}$ is the perturbative
decay rate in the vacuum.

The production of $\chi$ in the system introduced above has been also
studied in the context of parametric (or tachyonic) resonance
\cite{Kofman:1994rk, Kofman:1997yn, Dufaux:2006ee, Amin:2019qrx}.  In
particle production through the parametric resonance, the modes in
resonance bands are effectively produced.  In the present case, the
widths of the bands are determined by the following parameter:
\begin{align}
  q \equiv \frac{4A\bar{\phi}}{m_\phi^2}.
  \label{q-def}
\end{align}
As we will see below, the band width is $\sim qm_\phi$.  Hereafter,
for the comparison with the Boltzmann equation, we concentrate on the
case that the band widths are narrow, i.e.,
\begin{align}
  q \ll 1.
\end{align}
In the narrow resonance regime, $\chi$ is produced almost
``on-resonance,'' i.e., the momentum of $\chi$ produced from the
$\phi$ oscillation is $|\vec{k}|\simeq p_\chi$.  In the perturbative
description, effects which are higher order in $A$ are suppressed when
$q$ is small \cite{Matsumoto:2007rd}.

\section{Particle Production in Flat Spacetime}
\label{sec:flat}
\setcounter{equation}{0}

\subsection{Evolution equations from QFT}

In studying particle production in the QFT, an approach by employing
the Mathieu equation has been often used in the context of parametric
resonance.  In our analysis, however, we adopt a different approach,
which should be equivalent to the one with the Mathieu equation, by
deriving evolution equations for the distribution function.  An
advantage of our approach is that the equations directly describe the
evolution of the distribution function.  Because of this, we can
consider a possible relation between the evolution equation of the
distribution function in the QFT and the Boltzmann equation.

We first discuss particle production in the flat (i.e., Minkowski)
spacetime.  We decompose $\chi$ by using the creation and annihilation
operators, denoted as $a_{\vec{k}}$ and $a^\dagger_{\vec{k}}$,
respectively:
\begin{align}
  \chi(x)
  =\int{\frac{d^3k}{(2\pi)^3 \sqrt{2\omega_k}}}
  \left[
  e^{i \vec{k} \vec{x}} a_{\vec{k}} (t) +
  e^{-i \vec{k} \vec{x}} a^\dagger_{\vec{k}} (t)
  \right].
\end{align}
The creation and annihilation operators satisfy the following
commutation relation:
\begin{align}
  [ a_{\vec{k}}, a^{\dagger}_{\vec{k'}} ]
  =(2\pi)^3 \delta^{(3)} ( \vec{k}-\vec{k'} ).
  \label{CR}
\end{align}
With the creation and annihilation operators, the free part of the
Hamiltonian is given by
\begin{align}
  H_{\rm free} = \int \frac{d^3 k}{(2\pi)^3}
  \omega_{{k}} a_{\vec{k}}^\dagger a_{\vec{k}},
\end{align}
where
\begin{align}
  \omega_{k} \equiv \sqrt{|\vec{k}|^2 + m_\chi^2}.
\end{align}
In addition, the interaction Hamiltonian is expressed as
\begin{align}
  H_{\rm int}(t) = A \phi(t)
  \int \frac{d^3k}{(2\pi)^3 2 \omega_{{k}}}
  \left[ a^\dagger_{\vec{k}}(t) a_{\vec{k}}(t)
  + a_{\vec{k}}(t)a^\dagger_{\vec{k}}(t)
  + a_{\vec{k}}(t)a_{-\vec{k}}(t)
  +  a^\dagger_{\vec{k}}(t) a^\dagger_{-\vec{k}}(t) \right].
  \label{Hint}
\end{align}

In order to study the production of $\chi$, we introduce the
distribution function of $\chi$:
\begin{align}
  f_{\vec{k}} (t) \equiv
  \frac{1}{V} \langle a_{\vec{k}}^\dagger (t) a_{\vec{k}} (t) \rangle,
  \label{distfn_k}
\end{align}
where $V$ is the spatial volume, and the expectation value of the
operator ${\mathcal{O}}$ is defined by using a density matrix $\rho$
as
\begin{align}
  \langle{\mathcal{O}}\rangle\equiv\mbox{Tr}(\rho{\mathcal{O}}).
\end{align}  
Using the distribution function, the number density of $\chi$ is given
by
\begin{align}
\label{number1}
  n_\chi (t) = \int \frac{d^3k}{(2\pi)^3} f_{\vec{k}} (t).
\end{align}

In the following, we work in the interaction picture.  The time
dependence of the operator $\mathcal{O}$ is given by
\begin{align}
  \dot{\mathcal{O}} = -i [\mathcal{O}, H_{\rm free}],
\end{align}
where the ``dot'' denotes the derivative with respect to time, and
hence
\begin{align}
  \dot{a}_{\vec{k}} = -i \omega_k a_{\vec{k}}.
\end{align}
In addition, the density matrix evolves as
\begin{align}
  \dot{\rho} = i [\rho, H_{\rm int}].
\end{align}
Consequently, the evolution of the expectation value of the operator
$\mathcal{O}$ is governed by the following differential equation:
\begin{align}
  \frac{d}{dt} \langle \mathcal{O} \rangle =
  -i \langle [\mathcal{O}, H_{\rm int}] \rangle
  + \langle \dot{\mathcal{O}} \rangle.
\end{align}

In order to study the evolution of the distribution function, we also
introduce
\begin{align}
  g_{\vec{k}} \equiv
  \frac{1}{V} \langle a_{\vec{k}} a_{-\vec{k}} \rangle.
\end{align}
Then, the time derivatives of $f_{\vec{k}}$ and $g_{\vec{k}}$ are
given by
\begin{align}
  \dot{f}_{\vec{k}} = &\,
  i \frac{A \bar{\phi}}{\omega_k}
  \left( g_{\vec{k}} - g_{\vec{k}}^* \right) \cos m_\phi t,
  \label{fdot}
  \\
  \dot{g}_{\vec{k}} = &\,
  -i
  \left[ 2 \omega_k g_{\vec{k}} + \frac{A \bar{\phi}}{\omega_k}  
    \left( 1 + f_{\vec{k}} + f_{-\vec{k}} + 2 g_{\vec{k}} \right)
    \cos m_\phi t
    \right].
  \label{gdot}
\end{align}
We consider isotropic solutions, and the momentum dependences of
$f_{\vec{k}}$ and $g_{\vec{k}}$ are only via $k\equiv|\vec{k}|$; thus,
in the following, these functions are denoted as $f_k$ and $g_k$,
respectively.  We decompose the complex function $g_k$ as
\begin{align}
  g_k (t) = e^{-2i\omega_k t} \left[ \xi_k (t) + i \eta_k (t) \right],
\end{align}
where $\xi_k$ and $\eta_k$ are real functions.  Then, the evolution
equations become
\begin{align}
  \dot{f}_k = &\,
  \frac{A\bar{\phi}}{\omega_k}  \cos m_\phi t
  \left[ 2 \xi_k \sin 2 \omega_k t - 2 \eta_k \cos 2 \omega_k t \right],
  \label{fkdot}
  \\
  \dot{\xi}_k = &\,
  \frac{A\bar{\phi}}{\omega_k}  \cos m_\phi t
  \left[ (1 + 2f_k) \sin 2 \omega_k t + \eta_k \right],
  \label{xikdot}
  \\
  \dot{\eta}_k = &\,
  \frac{A\bar{\phi}}{\omega_k}  \cos m_\phi t
  \left[ - (1 + 2f_k) \cos 2 \omega_k t - \xi_k \right].
  \label{etakdot}
\end{align}
We note that the evolution equations above also apply to the case of
$q\gtrsim 1$, although we do not consider such a case in this paper.

Before solving these differential equations, we comment on the initial
condition.  Because we study the $\chi$ production from the $\phi$
oscillation, we consider the case that the $\chi$ sector is initially
at the ground state (vacuum), denoted as $|0\rangle$; thus, 
$\rho(0)=|0\rangle\langle 0|$.  Because $a_k|0\rangle=\langle
0|a_k^\dagger=0$, $f_k$ and $g_k$ should satisfy\footnote
{Our evolution equations are applicable to other types of initial
  conditions.}
\begin{align}
  f_k(0)=\xi_k(0)=\eta_k(0)=0.
  \label{initialcondition}
\end{align}

\begin{figure}
  \begin{center}  
    \includegraphics[width=0.9\textwidth]{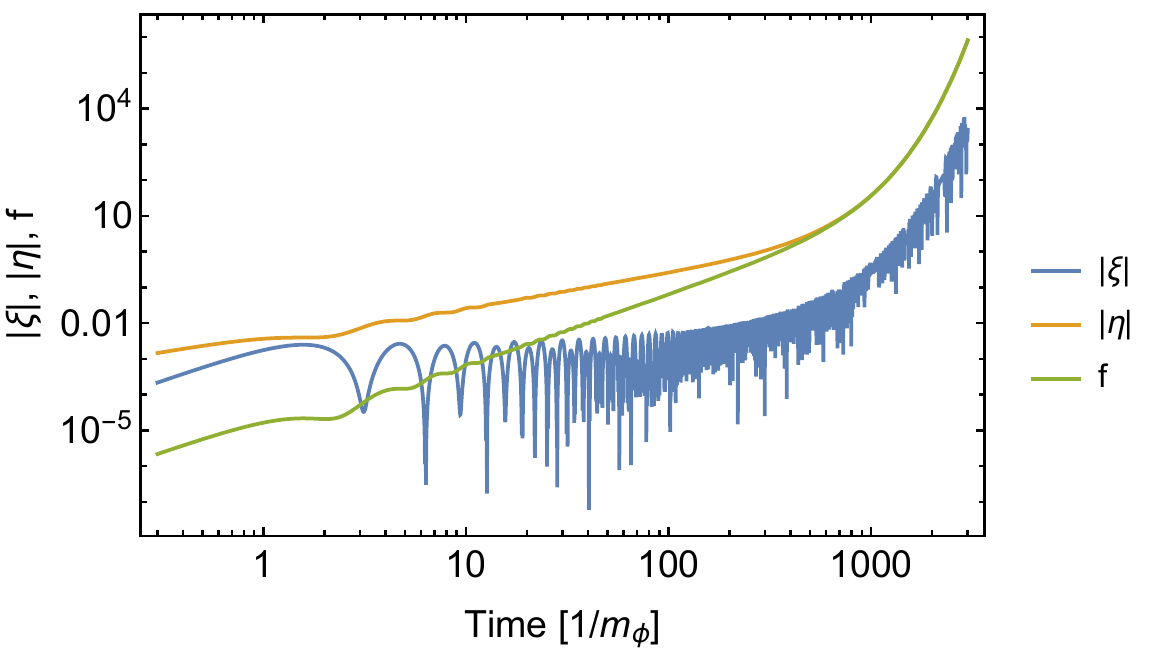}
  \end{center}
  \caption{$|\xi_k|$ (blue), $|\eta_k|$ (orange), and $f_k$ (green),
    as functions of time (in units of $m_\phi^{-1}$).  Here, we take
    $q=10^{-2}$, and $\omega=\frac{1}{2}m_\phi$.}
  \label{fig:fnevolve05m}
  \vspace{10mm}
  \begin{center}  
    \includegraphics[width=0.9\textwidth]{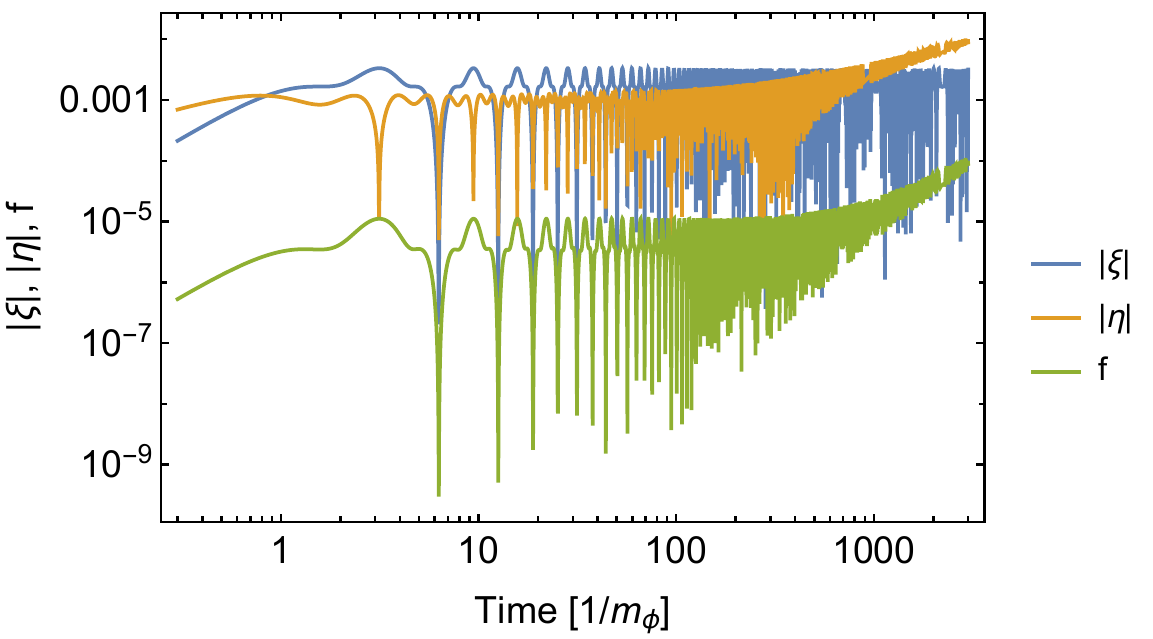}
  \end{center}
  \caption{Same as Fig.\ \ref{fig:fnevolve05m}, except for $\omega=m_\phi$.}
  \label{fig:fnevolve10m}
\end{figure}

We numerically solve Eqs.\ \eqref{fkdot} $-$ \eqref{etakdot} with the
initial condition given by Eq.\ \eqref{initialcondition}.  The
evolutions for $\omega_k=\frac{1}{2}m_\phi$ and $m_\phi$ are shown in
Figs.\ \ref{fig:fnevolve05m} and \ref{fig:fnevolve10m}, respectively, with
taking $q=10^{-2}$.  As one can see, $f_k$ monotonically increases when
$t\lesssim m_\phi$ irrespective of $\omega_k$ (as far as
$\omega_k\lesssim m_\phi$).  The behavior at $t\rightarrow\infty$
strongly depends on $\omega_k$.  For $\omega_k=\frac{1}{2}m_\phi$, the
functions $f_k$ and $\eta_k$ grow exponentially at
$t\rightarrow\infty$, while $\xi_k$ is oscillating.  On the contrary,
for $\omega=m_\phi$, $f_k$ does not show the growing behavior at
$t\rightarrow\infty$.

The behaviors of $f_k$, $\xi_k$, and $\eta_k$ at $t\ll m_\phi$ can be
obtained by expanding these functions as a power series of $t$.  With
$t$ being small enough, the effects of the oscillation of $\phi$ is
unimportant and we obtain
\begin{align}
  \left. f_k (t) \right|_{t\ll m_\phi^{-1}} \simeq &\,
  \frac{1}{4} q^2 m_\phi^2 t^2,
  \\
  \left. \xi_k (t) \right|_{t\ll m_\phi^{-1}} \simeq &\,
  \frac{1}{4} q m_\phi^2 t^2,
  \\
  \left. \eta_k (t) \right|_{t\ll m_\phi^{-1}} \simeq &\,
  - \frac{1}{2} q m_\phi t.
\end{align}

We can also understand the exponential growths of the modes with
$\omega_k\sim\frac{1}{2}m_\phi$ at $t\rightarrow\infty$. We first
adopt the following ansatz:
\begin{align}
  \left. f_k (t) \right|_{t\gtrsim (qm_\phi)^{-1}} \simeq 
  \alpha_k e^{\lambda_k t},
  \label{expansatz}
\end{align}
where $\alpha_k$ and $\lambda_k$ are positive constants.  Substituting
the above ansatz, Eq.\ \eqref{gdot} becomes
\begin{align}
  g_k (t) \simeq
  \frac{2A\bar{\phi}}{\omega_k}
  \frac{1}{m_\phi^2 - (2\omega_k-i\lambda_k)^2}
  \left[ (2\omega_k -i \lambda_k)\cos m_\phi t -i m_\phi \sin m_\phi t \right]
  f_k (t),
  \label{g(f>>1)}
\end{align}
where we neglect terms which are not exponentially growing, and
\begin{align}
  \mbox{Im} g_k (t) \simeq &\,
  - \frac{A\bar{\phi}}{\omega_k}
  \frac{1}{(2\omega_k - m_\phi)^2 + \lambda_k^2}
  \left[ \lambda_k \cos m_\phi t - (2\omega_k - m_\phi) \sin m_\phi t \right]
  f_k (t)
  \nonumber \\ &\,
  - \frac{A\bar{\phi}}{\omega_k}
  \frac{1}{(2\omega_k + m_\phi)^2 + \lambda_k^2}
  \left[ \lambda_k \cos m_\phi t + (2\omega_k + m_\phi) \sin m_\phi t \right]
  f_k (t).
  \label{Im(g)}
\end{align}
For the study of the modes with $\omega_k\sim\frac{1}{2}m_\phi$, we
neglect the second term of the right-hand side of the above equation
because it is sub-dominant.  Combining Eqs.\ \eqref{fdot} and
\eqref{Im(g)}, we obtain
\begin{align}
  \frac{\dot{f}_k}{f_k} \simeq
  2 \left( \frac{A\bar{\phi}}{\omega_k} \right)^2
  \frac{1}{(2\omega_k - m_\phi)^2 + \lambda_k^2}
  \cos m_\phi t
    \left[ \lambda_k \cos m_\phi t - (2\omega_k - m_\phi) \sin m_\phi t \right].
\end{align}
We neglect terms which are oscillating with the timescale of $\sim
m_\phi^{-1}$ because we are not interested in terms with rapid
oscillations.  We define the time average of the function $F(t)$ as
\begin{align}
  \langle F \rangle_t
  \equiv
  \frac{1}{t_*} \int_t^{t+t_*} dt' F (t'),
\end{align}
where the timescale $t_*$ is taken to be $m_\phi^{-1}\ll t_*\ll
(qm_\phi)^{-1}$.  Using the relation $\langle\cos^2 m_\phi
t\rangle_t\simeq\frac{1}{2}$ and $\langle\sin 2m_\phi t\rangle_t\simeq
0$, we can see that Eqs.\ \eqref{fdot} and \eqref{gdot} are satisfied
with the ansatz given in Eq.\ \eqref{expansatz} (neglecting rapidly
oscillating terms and terms which do not grow exponentially) if
\begin{align}
  \lambda_k = 
  \sqrt{ \frac{1}{4} \left( \frac{m_\phi}{2\omega_k} \right)^2 q^2 m_\phi^2
    - (2\omega_k - m_\phi)^2}.
  \label{lambda_k(exact)}
\end{align}
The growth rate $\lambda_k$ should be real, which determines the
resonance band.  We can see that the width of the resonance band is of
$O(qm_\phi)$.  Then, keeping the leading term in $q$, we may use
\begin{align}
  \lambda_k \simeq
  \sqrt{ \frac{1}{4} q^2 m_\phi^2 - (2\omega_k - m_\phi)^2},
  \label{lambda_k(approx)}
\end{align}
and the resonance band is given by
$\omega_-\lesssim\omega_k\lesssim\omega_+$, where
\begin{align}
  \omega_\pm \equiv \frac{1}{2} m_\phi
  \left( 1 \pm \frac{1}{2} q \right).
\end{align}
Notice that the resonance band given above corresponds to the first
resonance band in the study of the parametric resonance, and that the
growth rate given in Eq.\ \eqref{lambda_k(approx)} is consistent with
that given in \cite{Yoshimura:1995gc, Kofman:1997yn, Dufaux:2006ee}.

\begin{table}[t]
  \begin{center}
    \begin{tabular}{rrr}
      \hline\hline
      $z$ & Numerical & Eq.\ \eqref{expansatz}
      \\
      \hline
      $-0.4$ & $7.3\times 10^{11}$ & $6.2\times 10^{11}$
      \\
      $-0.2$ & $8.9\times 10^{17}$ & $8.6\times 10^{17}$
      \\
      $0$    & $3.5\times 10^{19}$ & $3.5\times 10^{19}$
      \\
      $0.2$  & $7.7\times 10^{17}$ & $7.8\times 10^{17}$
      \\
      $0.4$  & $4.7\times 10^{11}$ & $4.6\times 10^{11}$
      \\
      \hline\hline
    \end{tabular}
    \caption{The ratio
      $f_k(t=100(qm_\phi)^{-1})/f_k(t=10(qm_\phi)^{-1})$, based on the
      numerical calculation (Numerical) and the approximation adopting
      the ansatz of the exponential increase (Eq.\ \eqref{expansatz})
      with the growth rate given in Eq.\ \eqref{lambda_k(exact)}.  We
      take $q=10^{-2}$, and several values of $\omega_k$ (parameterized
      as $\omega_k=\frac{1}{2}m_\phi (1+z q)$).}
    \label{table:expgrowth}
  \end{center}
\end{table}

In order to check the validity of the ansatz of the exponential
increase given in Eq.\ \eqref{expansatz}, we compare the value of
$f_k(t=100(qm_\phi)^{-1})/f_k(t=10(qm_\phi)^{-1})$ obtained from the
numerical calculation and that predicted by Eq.\ \eqref{expansatz}
(with the growth rate given in Eq.\ \eqref{lambda_k(exact)}).  The
results with taking $q=10^{-2}$ are shown in Table
\ref{table:expgrowth} for several values of $\omega_k$ in the
resonance band.  Notice that $f_k(t=10(qm_\phi)^{-1})$ is of $O(10)$
for the frequencies considered in Table \ref{table:expgrowth}, and
hence the effect of stimulated emission is already important when
$t=10(qm_\phi)^{-1}$.  We can see that the occupation number in the
resonance band is well described by the exponential increase with the
growth rate given above once the effect of the stimulated emission
becomes effective.

\subsection{QFT and ordinary Boltzmann equation}

The relation between the analysis so far and that based on the
conventional Boltzmann equation is interesting. Sometimes particle
production from the oscillating scalar field $\phi$ is studied by
using Boltzmann equation.  $\phi$ can be regarded as a coherent state
of a non-relativistic scalar field and the production of $\chi$ may be
regarded as the decay of $\phi$.  Then, the collision term in the
Boltzmann equation is given by (see, for example, \cite{Kolb:1990vq,
  Mukhanov:2005sc})
\begin{align}
  \dot{f}_k^{\rm (coll)} (t)
  = 
  \frac{4\pi^2 n_\phi \Gamma_{\phi\rightarrow\chi\chi}^{(0)}}{p_\chi^2}
  \left[ (1 + f_k)^2 - f_k^2 \right]
  \delta( k - p_\chi ),
  \label{collisionterm}
\end{align}
where the first and the second terms in the square bracket describe
the effects of the decay and the inverse decay of $\phi$, respectively.
In the following, we consider how the collision term can be related to
the argument based on the QFT.  We treat the cases of $t\lesssim
(qm_\phi)^{-1}$ and $t\gtrsim (qm_\phi)^{-1}$ separately.

When $t\lesssim (qm_\phi)^{-1}$, $f_k$ is smaller than $1$ and the
solution of Eq.\ \eqref{gdot} is approximately given by
\begin{align}
  g_k (t) \simeq
  -\frac{1}{4\omega_k^2 - m_\phi^2}
  \frac{A\bar{\phi}}{\omega_k}
  \left(
  -2 c \omega_k e^{-2i\omega_k t}
  + 2 \omega_k \cos m_\phi t - i m_\phi \sin m_\phi t
  \right),
  \label{gk(approx)}
\end{align}
where $c$ is a constant.  Here, we have neglected the last term of the
right-hand side of Eq.\ \eqref{gdot} because it is of $O(q)$ and is
sub-dominant relative to the first term.\footnote
{The homogeneous solution of Eq.\ \eqref{gdot}, i.e., the solution
  with taking $(1+2f_k)\rightarrow 0$, is given by
  \begin{align*}
    g_k^{\rm (homogenious)} (t) =
    c' \exp
    \left( -2i\omega_k t - i \frac{A\bar{\phi}}{m_\phi \omega_k} \sin m_\phi t
    \right),
  \end{align*}
  with $c'$ being a constant.  In Eq.\ \eqref{gk(approx)}, the second
  term in the bracket is neglected because it gives a higher order
  contribution in terms of $q$.}
Because $g_k(0)=0$ and also because $g_k$ is non-singular at
$\omega_k\rightarrow\frac{1}{2}m_\phi$, $c=1$ and hence $\dot{f}_k$
becomes
\begin{align}
  \left. \dot{f}_k \right|_{t\lesssim (qm_\phi)^{-1}}
  \simeq
  \frac{\pi}{4} \left( \frac{A\bar{\phi}}{\omega_k} \right)^2
  \gamma (\omega_k; t), 
\end{align}
where
\begin{align}
  \gamma (\omega; t) \equiv
  \frac{8}{\pi (4\omega^2 - m_\phi^2)} \cos m_\phi t
  \left( 2\omega \sin 2 \omega t - m_\phi \sin m_\phi t \right).
\end{align}
In the above expression, we neglect the terms which are rapidly
oscillating with the timescale of $\sim m_\phi^{-1}$ (even in the
limit of $\omega_k\rightarrow\frac{1}{2}m_\phi$).  Notice that the
function $\gamma (\omega; t)$ has the following property:
\begin{align}
  \int_0^\infty d\omega
  \gamma (\omega; t) = 1 + \cos 2 m_\phi t.
  \label{intgamma}
\end{align}
We consider the time averaged value of $\dot{f}_k$ and neglect terms
which are rapidly oscillating:
\begin{align}
  \left. \langle \dot{f}_k \rangle_t \right|_{t\lesssim (qm_\phi)^{-1}}
  \simeq
  \frac{\pi}{4} \left( \frac{A\bar{\phi}}{\omega_k} \right)^2
  \Delta (\omega_k; t_*),
\end{align}
where
\begin{align}
  \Delta (\omega; t_*) \equiv
  \frac{1}{t_*}
  \int_0^{t_*} dt \gamma (\omega; t).
\end{align}
Using Eq.\ \eqref{intgamma}, we obtain
\begin{align}
  \lim_{t_*\rightarrow\infty}
  \int_0^\infty d\omega \Delta (\omega; t_*)
  = \lim_{t_*\rightarrow\infty}
  \frac{1}{t_*} \int_t^{t+t_*} dt' \left( 1 + \cos 2 m_\phi t' \right)
  = 1.
\end{align}
In addition,
\begin{align}
  \left. \lim_{t_*\rightarrow\infty} \Delta (\omega; t_*) \right|_{2|\omega|\neq m_\phi}
  = 0.
\end{align}
Thus, assuming that $t_*$ can be chosen to be large enough, we may
approximate
\begin{align}
  \Delta (\omega; t_*) \rightarrow
  \delta (\omega - \mbox{$\frac{1}{2}$} m_\phi),
\end{align}
and obtain
\begin{align}
  \left. \langle \dot{f}_k \rangle_t \right|_{t\lesssim (qm_\phi)^{-1}}
  \simeq
  \frac{4\pi^2 n_\phi \Gamma_{\phi\rightarrow\chi\chi}^{(0)}}{p_\chi^2}
  \delta( k - p_\chi ).
  \label{fcoll_1}
\end{align}

When $t\gtrsim (qm_\phi)^{-1}$, $f_k$ in the resonance band shows the
exponential growth and is much larger than $1$, while $g_k$ behaves as
Eq.\ \eqref{g(f>>1)}.
In such a case, the conventional Boltzmann equation is obtained if
one performs the following replacement in Eq.\ \eqref{g(f>>1)}:
\begin{align}
  \frac{1}{m_\phi^2 - (2\omega_k-i\lambda_k)^2}
  \rightarrow
  -i \frac{\pi}{2m_\phi} \delta (2\omega_k - m_\phi)
  + (\mbox{irrelevant}).
  \label{replace}
\end{align}
Here, the relation $\frac{1}{x-i0}=P(\frac{1}{x})+i\pi\delta (x)$
(with $P$ being the principal value), as well as the smallness of
$\lambda_k$ relative to $m_\phi$, are used.  However, we should note
that this replacement is allowed if $\lambda_k$ does not depend much
on $\omega_k$ and also if $f_k(t)$ is insensitive to $k$ around the pole.
These requirements may not be satisfied in the case of our interest
(see Eq.\ \eqref{lambda_k(approx)}), which may result in the
non-agreements between the result of the QFT calculation and that of the
Boltzmann equation.  Substituting Eq.\ \eqref{g(f>>1)} (with using the
above replacement) into Eq.\ \eqref{fdot}, we obtain
\begin{align}
  \left. \dot{f}_k \right|_{t\gtrsim (qm_\phi)^{-1}} \simeq
  8\pi \left( \frac{A\bar{\phi}}{m_\phi} \right)^2
  \cos^2 m_\phi t
  f_k (t) \delta (2\omega_k - m_\phi).
\end{align}
Neglecting the rapidly oscillating term with taking the time average,
\begin{align}
  \left. \langle \dot{f}_k \rangle_t \right|_{t\gtrsim (qm_\phi)^{-1}}
  \simeq
  \frac{8\pi^2 n_\phi \Gamma_{\phi\rightarrow\chi\chi}^{(0)}}{p_\chi^2}
  f_k (t)
  \delta( k - p_\chi ).
  \label{fcoll_2}
\end{align}

Combining the results for $t\lesssim (qm_\phi)^{-1}$ and $t\gtrsim
(qm_\phi)^{-1}$, which are given in Eqs.\ \eqref{fcoll_1} and
\eqref{fcoll_2}, respectively,
and estimating the collision term as
\begin{align}
  \dot{f}_k^{\rm (coll)} (t) \simeq \langle \dot{f}_k (k) \rangle_t,
\end{align}
we find that the collision term given in \eqref{collisionterm} well
describes the evolution of $f_k$ if the assumptions and approximations
adopted in the above argument are valid. 

\begin{figure}[t]
  \begin{center}  
    \includegraphics[width=0.9\textwidth]{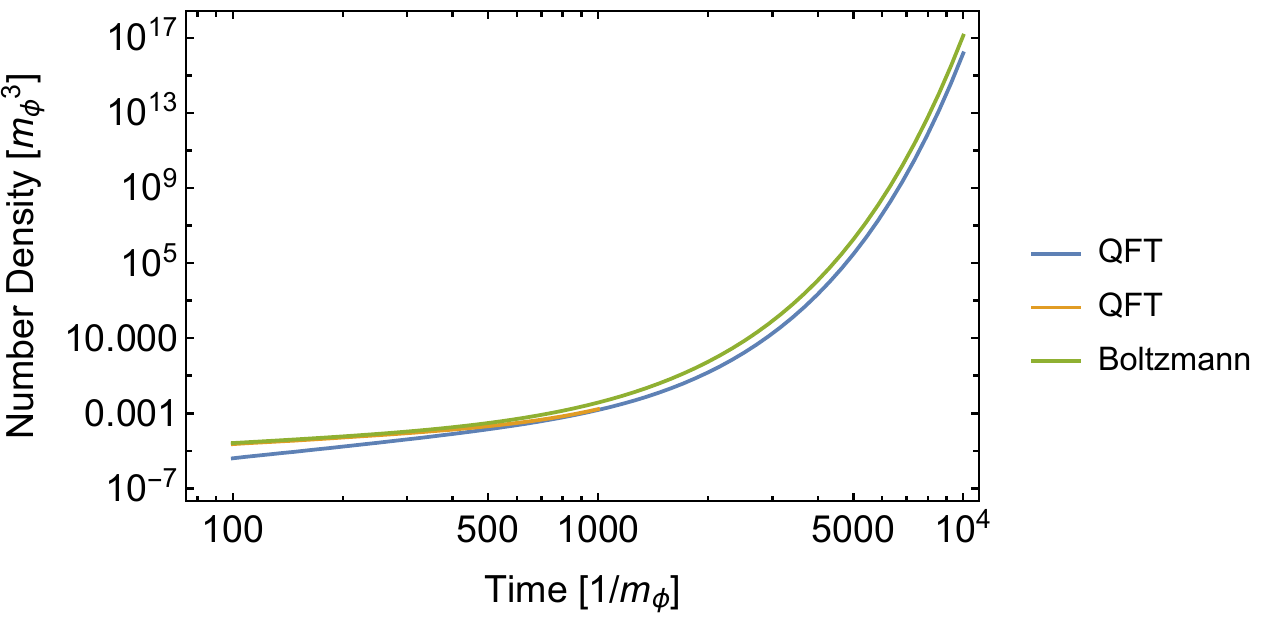}
  \end{center}
  \caption{Evolution of $n_\chi^{\rm (QFT)}$ (in units of
    $m_\phi^3$) for $q=10^{-2}$, adopting the integration regions
    of $|\omega-\frac{1}{2}m_\phi|\leq \frac{1}{4}qm_\phi$ and
    $\frac{5}{2}qm_\phi$ (blue and orange, respectively).  We also show the
    evolution of $n_\chi^{\rm (Boltzmann)}$ (green).}
  \label{fig:ndmtdep}
\end{figure}

In order to see the validity of the collision term from the QFT point
of view, we calculate the number density of $\chi$ with two different
methods, taking $m_\chi=0$ (and hence $\omega_k=k$) and $q=10^{-2}$.
\begin{itemize}
\item First, we use the QFT approach; the number density based on the
  QFT is denoted as $n_\chi^{\rm (QFT)}$.  Eq.\ \eqref{number1} gives
  \begin{align}
    n_\chi^{\rm (QFT)} (t) = \frac{1}{2\pi^2} \int dk k^2
    f_k (t).
  \end{align}
  When $t\gtrsim (qm_\phi)^{-1}$, $f_k(t)$ is sharply peaked at
  $k\sim\frac{1}{2}m_\phi$, and $n_\chi^{\rm (QFT)}$ is insensitive to
  the integration region.  Here, unless otherwise mentioned, we choose
  the integration region to be within the resonance band:
  $|k-\frac{1}{2}m_\phi|\leq \frac{1}{4}qm_\phi$.
\item Second, we use the Boltzmann equation; the number density based
  on the Boltzmann equation is denoted as $n_\chi^{\rm (Boltzmann)}$.
  Solving $\dot{f}_k=\dot{f}_k^{\rm (coll)}$, we obtain
  \begin{align}
    n_\chi^{\rm (Boltzmann)} (t) = 2 \Gamma_\phi^{(0)} n_\phi \int_0^t dt'
    \left( 1 + 2f_{p_\chi} (t') \right).
    \label{n^B}
  \end{align}
  Notice that, in the above expression, the distribution function in
  the parenthesis is that for $k=p_\chi$ and is obtained by
  solving Eqs.\ \eqref{fdot} and \eqref{gdot}.
\end{itemize}
The time dependences of $n_\chi^{\rm (QFT)}$ and $n_\chi^{\rm
  (Boltzmann)}$ are shown in Fig.\ \ref{fig:ndmtdep}.  We find that
the behaviors of $n_\chi^{\rm (QFT)}$ and $n_\chi^{\rm (Boltzmann)}$
are similar, although $n_\chi^{\rm (QFT)} (t)$ is smaller.  For $t\sim
(qm_\phi)^{-1}$, the width of the peak of $f_k$ around
$k=\frac{1}{2}m_\phi$ is significantly larger than
$\frac{1}{4}qm_\phi$, and hence $n_\chi^{\rm (QFT)} (t)$
underestimates the number density with our choice of the integration
region.  If we expand the integration region as
$|k-\frac{1}{2}m_\phi|\leq \frac{5}{2}qm_\phi$, for example, the
agreement for $t\sim (qm_\phi)^{-1}$ becomes much better; difference
between $n_\chi^{\rm (QFT)}$ and $n_\chi^{\rm (Boltzmann)}$ is $\sim
10\ \%$ with such a choice of the integration region.  In
Fig.\ \ref{fig:ndmtdep}, we also show $n_\chi^{\rm (QFT)}$ with
adopting the integration region of $|k-\frac{1}{2}m_\phi|\leq
\frac{5}{2}qm_\phi$.  (We show such a result only for relatively small
$t$ because the calculation of $n_\chi^{\rm (QFT)}$ with larger $t$
requires a computational cost.)  For $t\gg (qm_\phi)^{-1}$, on the
contrary, the peak width of $f_k$ is smaller than the width of the
resonance band (see below).  For the replacement given in
Eq.\ \eqref{replace}, it is required that the function $f_k$ is well
approximated by $f_{p_\chi}$ for $|k-\frac{1}{2}m_\phi|\lesssim
\lambda_k$ and that the $\omega_k$ dependence of $\lambda_k$ is
unimportant.  These cannot be the case in particular when $t\gg
(qm_\phi)^{-1}$.  Consequently, $n_\chi^{\rm (Boltzmann)}$ becomes
larger than $n_\chi^{\rm (QFT)}$, as shown in Fig.\ \ref{fig:ndmtdep}.

We can understand the asymptotic behaviors of $n_\chi^{\rm (QFT)}$ and
$n_\chi^{\rm (Boltzmann)}$ for $t\gg (qm_\phi)^{-1}$ as follows.  At
$t\gg (qm_\phi)^{-1}$, as we see below, $f_k$ is sharply peaked at
$\omega_k=\frac{1}{2}m_\phi$.  We expand $\lambda_k$ around the peak
and obtain
\begin{align}
  f_{k}(t)\simeq \alpha_k e^{\frac{q m_\phi t}{2}}
  e^{-\frac{(2 \omega_k -m_\phi)^2}{q m_\phi/t}}.
\end{align}
Because of the second exponential factor, the width of $f_k(t)$ gets
smaller as $t$ increases.
 By using the above expression with
neglecting $m_\chi$, $n_\chi^{\rm (QFT)}(t)$ at $t\gg (qm_\phi)^{-1}$
is estimated as
\begin{align}
  n_\chi^{\rm (QFT)}(t) \sim
  \frac{1}{16\pi} (\pi q m_\phi t)^{-1/2} q m_\phi^3 f_{p_\chi}.
\end{align}
We can see that the above expression is in a good agreement with the
numerical result.  In addition,
\begin{align}
  n_\chi^{\rm (Boltzmann)}(t) \sim
  \frac{1}{32\pi} q m_\phi^3 f_{p_\chi}.
\end{align}
Thus, when $t\gg (qm_\phi)^{-1}$, $n_\chi^{\rm (Boltzmann)}(t)$
becomes larger than $n_\chi^{\rm (QFT)}(t)$. 

One may wonder if we can introduce an ``averaged'' occupation number
(or growth rate) in the resonance band to make two approaches
consistent.  However, as we will see in the next section, the
discrepancy in the case with the cosmic expansion cannot be solved
with such a prescription.  (In addition, in Appendix
\ref{app:boltzmann}, we give a consideration about the relation
between the QFT and Boltzmann equation.)

Before closing this section, we comment on the back reaction.  In our
analysis, the effects of the back reaction are neglected; we assume
that the amplitude of the $\phi$ oscillation does not depend on time
and that the motion of $\phi$ is well described by
Eq.\ \eqref{phioscillation}.  This is the case when the energy density
transferred to the $\chi$ sector is smaller than the initial energy
density in the $\phi$ sector.  Let us denote the typical value of the
occupation number of the modes in the resonance band as $f^{\rm
  (res)}$; conservatively, we may take $f^{\rm (res)}\sim
f_{k=p_\chi}$.  Then, by using the fact that the width of the
resonance band is $\sim qm_\phi$, the energy density in the $\chi$
sector is estimated as $\rho_\chi\sim f^{\rm (res)}qm_\phi^4$.
Requiring that $\rho_\chi$ is smaller than the initial energy density
of $\phi$, we obtain
\begin{align}
  f^{\rm (res)} q m_\phi^2 \ll \bar{\phi}^2,
  \label{Econservation}
\end{align}
and, equivalently, $f^{\rm (res)}A\ll\bar{\phi}$.  The above
constraint can be satisfied for any value of $q$ by taking large
enough $\bar{\phi}$ (and small enough $A$).  Our results are
applicable to the parameter region consistent with the above
constraint.  We may also define the effective decay rate
$\Gamma_{\phi\rightarrow\chi\chi}^{\rm (eff)}$ as the inverse of the
timescale with which a single $\phi$ becomes a pair of $\chi$; we can
estimate $\Gamma_{\phi\rightarrow\chi\chi}^{\rm (eff)}\sim f^{\rm
  (res)}\Gamma_{\phi\rightarrow\chi\chi}^{(0)}$.  With the constraint
\eqref{Econservation} being satisfied,
$\Gamma_{\phi\rightarrow\chi\chi}^{\rm (eff)}$ is always smaller than
$qm_\phi$ as far as $q\ll 1$. Another back reaction may be due to the
scattering process like $\chi\phi\rightarrow\chi\phi$ (with $\phi$ in
the initial state being that in the coherent oscillation).  Such a
scattering process may remove $\chi$ from the resonance band.  Effects
of the scattering processes are not taken into account in our analysis
because $\phi$ is treated as a classical field.  One can estimate the
interaction rate as $\Gamma^{(\rm scat)}\sim q^4
\frac{m_\phi^3}{\bar{\phi}^2}$ which can be neglected compared with
the growth rate which is of $O(q m_\phi)$.

\section{Particle Production with Cosmic Expansion}
\label{sec:cosmo}
\setcounter{equation}{0}

In this section, we study particle production taking into account
the effects of the cosmic expansion.  We show that the analysis based
on the Boltzmann equation may result in a significant overestimation
of the occupation number of $\chi$ compared to the analysis based on
the QFT.

\subsection{Particle production with cosmic expansion from QFT}

With the cosmic expansion, the
momentum of each mode redshifts.  Thus, if we consider a mode which
has a frequency larger than $\omega_+$ in an early epoch, it enters
the resonance band as the Universe expands, then the frequency becomes
smaller than $\omega_-$ and the mode exits the resonance band.  The
occupation number may exponentially increase when the mode is in the
resonance band, as we see below.

Here, we are particularly interested in the behavior of the occupation
number when the mode is around the resonance band.  The timescale
of the mode to go through the resonance band is $\sim qH^{-1}$, where
$H$ is the expansion rate:
\begin{align}
  H \equiv \frac{\dot{a}}{a},
\end{align}
with $a$ being the scale factor.  The timescale of the evolution of
the occupation number is much shorter than the timescale of the cosmic
expansion because $q\ll 1$ and hence the change of $H$ is unimportant.
In the following analysis, we neglect the time dependence of $H$ and
take
\begin{align}
  a(t) = a_0 e^{Ht},
\end{align}
with $a_0$ being a constant. Due to the same reason, again we treat
$\bar{\phi}$ as constant. 

Because of the hierarchy between the timescales of the cosmic
expansion and the evolutions of $f_k$ and $g_k$, for the timescale
shorter than $H^{-1}$, evolutions of $f_k$ and $g_k$ are expected to
be the same as those in the flat spacetime.  Then, $f_k$ and $g_k$
obey
\begin{align}
  \dot{f}_k = &\,
  k H \frac{\partial f_k}{\partial k}
  + i \frac{A \bar{\phi}}{\omega_k}
  \left( g_k - g_k^* \right) \cos m_\phi t,
  \label{fdot_c}
  \\
  \dot{g}_k = &\,
  k H \frac{\partial g_k}{\partial k}
  -i
  \left[ 2 \omega_k g_k + \frac{A \bar{\phi}}{\omega_k}  
    \left( 1 + 2f_k + 2 g_k \right)
    \cos m_\phi t
    \right],
  \label{gdot_c}
\end{align}
where the first terms of the right-hand sides of the above equations
describe the effect of redshift.  We can simplify solve the above equations
by introducing functions for a fixed comoving momentum:
\begin{align}
  f_\star (t) \equiv f_{k=K(t)} (t),~~~
  g_\star (t) \equiv g_{k=K(t)} (t),
\end{align}
where $K(t)$ is the physical momentum for the given comoving momentum
$\hat{k}$ (which is independent of time):
\begin{align}
  K (t) \equiv \frac{a_0}{a(t)} \hat{k}.
\end{align}
We decompose the function $g_\star$ using real functions $\xi_\star$
and $\eta_\star$ as
\begin{align}
  g_\star (t) = e^{-2i\Theta (t)}
  \left[ \xi_\star (t) + i \eta_\star (t) \right],
\end{align}
where
\begin{align}
  \Theta (t) \equiv
  \int^t dt' \Omega (t'),
\end{align}
with
\begin{align}
  \Omega (t) \equiv \omega_{K(t)} = 
  \sqrt{ K^2(t) + m_\chi^2}.
\end{align}
Then, Eqs.\ \eqref{fdot_c} and \eqref{gdot_c} become
\begin{align}
  \dot{f}_\star = &\,
  \frac{A\bar{\phi}}{\Omega}  \cos m_\phi t
  \left[ 2 \xi_\star \sin 2 \Theta - 2 \eta_\star \cos 2 \Theta \right],
  \label{f*dot}
  \\
  \dot{\xi}_\star = &\,
  \frac{A\bar{\phi}}{\Omega}  \cos m_\phi t
  \left[ (1 + 2f_\star) \sin 2 \Theta + \eta_\star \right],
  \\
  \dot{\eta}_\star = &\,
  \frac{A\bar{\phi}}{\Omega}  \cos m_\phi t
  \left[ - (1 + 2f_\star) \cos 2 \Theta - \xi_\star \right].
  \label{eta*dot}
\end{align}

\begin{figure}
  \begin{center}
    \includegraphics[width=0.8\textwidth]{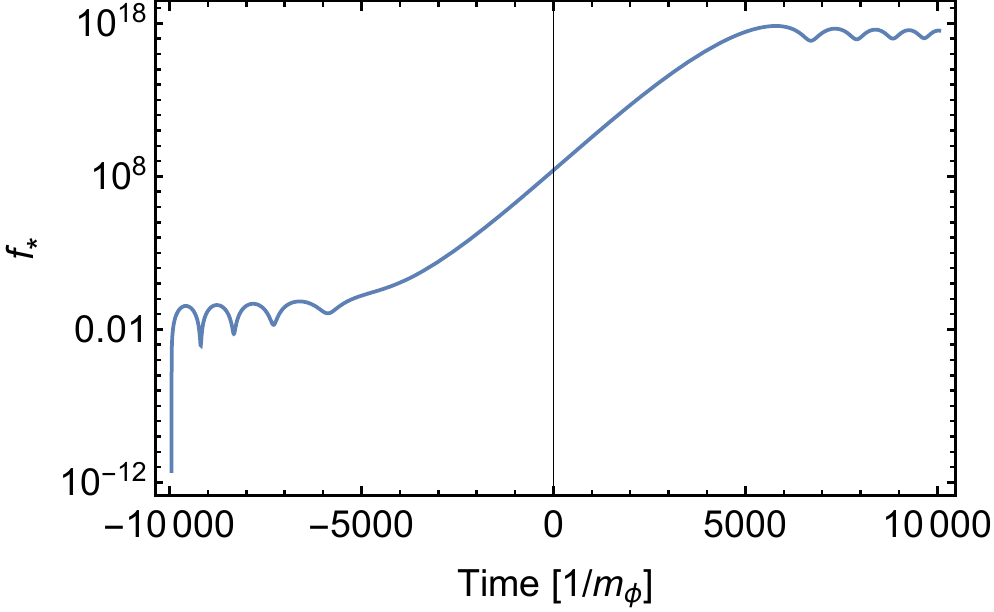}
  \end{center}
  \caption{Evolution of $f_\star$, taking $q=10^{-2}$ and
    $H=10^{-2}q^2\ m_\phi$ (for which $t_\pm\simeq \mp 5000\ m_\phi^{-1}$).}
  \label{fig:fcosmoh1em2}
  \vspace{10mm}
  \begin{center}  
    \includegraphics[width=0.8\textwidth]{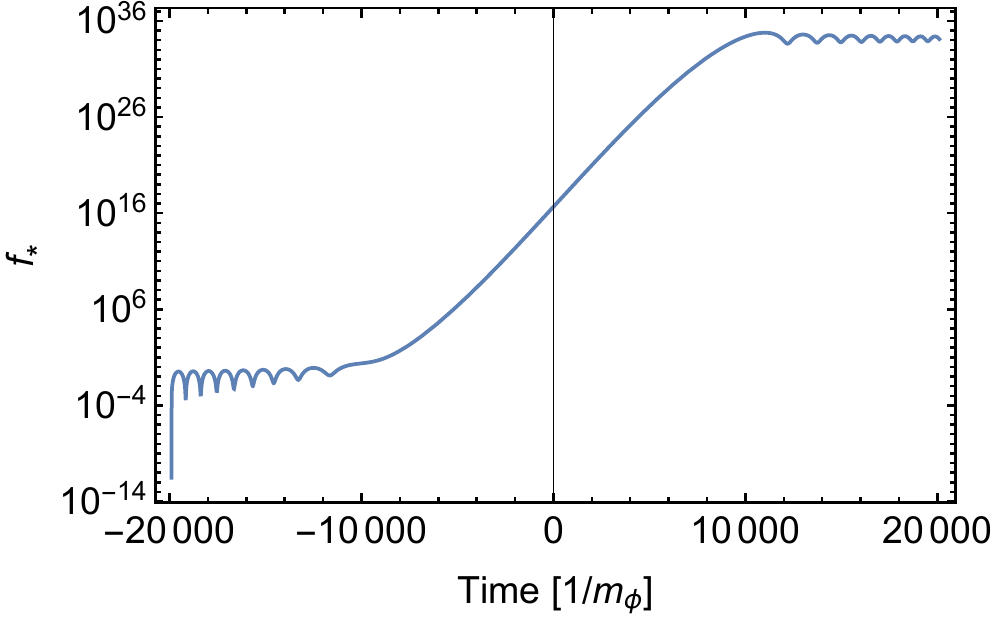}
  \end{center}
  \caption{Evolution of $f_\star$, taking $q=10^{-2}$ and $H=5\times
    10^{-3}\ q^2m_\phi$ (for which $t_\pm\simeq \mp 10000\ m_\phi^{-1}$).}
  \label{fig:fcosmoh5em3}
\end{figure}

We numerically solve Eqs.\ \eqref{f*dot} $-$ \eqref{eta*dot}.  We take
$m_\chi=0$ for simplicity, and 
\begin{align}
  a_0 \hat{k}=\frac{1}{2}m_\phi,
  \label{a0khat}
\end{align}
i.e., $\Omega (0)=\frac{1}{2}m_\phi$.  We choose $t_{\rm i}$ and
$t_{\rm f}$ such that $\Omega (t_{\rm i})\geq\omega_+$ and $\Omega
(t_{\rm f})\leq\omega_-$, and follow the evolutions of $f_\star$ and
$g_\star$ from $t=t_{\rm i}$ to $t=t_f$; here, we take
\begin{align}
  t_{\rm i} = -\frac{1}{H} \ln (1+q),~~~
  t_{\rm f} = -\frac{1}{H} \ln (1-q),
  \label{titf}
\end{align}
and hence
\begin{align}
  \Omega (t_{\rm i}) = \frac{1}{2} m_\phi (1+q),~~~
  \Omega (t_{\rm f}) = \frac{1}{2} m_\phi (1-q).
\end{align}
The cosmic time at which the mode enters (exits) the resonance band is
denoted as $t_+$ ($t_-$):
\begin{align}
  t_\pm \equiv
  - \frac{1}{H} \ln \left( 1 \pm \frac{1}{2} q \right),
\end{align}
i.e., $\Omega(t_\pm)=\omega_\pm$.  We impose the following initial
condition at $t_{\rm i}$:
\begin{align}
  f_\star (t_{\rm i}) = g_\star (t_{\rm i}) = 0.
  \label{initialcond}
\end{align}
We show the behavior of $f_\star$ for $q=10^{-2}$ in
Figs.\ \ref{fig:fcosmoh1em2} and \ref{fig:fcosmoh5em3}, in which we
take $H=10^{-2}q^2m_\phi$ and $H=5\times 10^{-3}q^2m_\phi$,
respectively.  We can observe exponential growths of $f_\star$ when
$\omega_-\lesssim\Omega(t)\lesssim\omega_+$, while there is no
significant increase of the occupation number when the mode is outside
of the resonance band.

For the case that $f_\star(t_-)\gg 1$, the total amount of the
increase of $f_\star$ can be estimated by using the growth rate
$\lambda_k$ given in the previous section.  Because the enhancement of
$f_\star$ occurs when the mode is in the resonance band, we
concentrate on the epoch of $t_+\leq t\leq t_-$; during such an epoch,
the occupation number is expected to evolve as $\dot{f}_k\simeq
kH(\partial f_k/\partial k)+\lambda_k f_k$.  Then, adopting the
growth rate given in Eq.\ \eqref{lambda_k(approx)},
$\dot{f}_\star$ is given by
\begin{align}
  \dot{f}_\star \simeq
  \sqrt{\frac{1}{4} q^2 m_\phi^2 - 
    \left( 2 \Omega - m_\phi \right)^2 }
  f_\star,
\end{align}
which results in
\begin{align}
  \frac{f_\star (t_-)}{f_\star (t_+)} \simeq
  \exp
  \left[ \int_{t_+}^{t_-} dt'  
    \sqrt{\frac{1}{4} q^2 m_\phi^2 - 
      \left( 2 \Omega - m_\phi \right)^2 }
      \right].
\end{align}
For the case that the mass of $\chi$ is negligible, the above ratio
is estimated as \cite{Shtanov:1994ce,  Kasuya:1996np, Kofman:1997yn}
\begin{align}
  \frac{f_\star (t_-)}{f_\star (t_+)} \simeq
  \exp \left( \frac{\pi q^2 m_\phi}{8H} \right).
  \label{fstarratio}
\end{align}
In order for a significant enhancement of the occupation number, $H$
should be smaller than $q^2 m_\phi$.

To show the validity of the above estimation, we calculate
the ratio $f_\star(t_-)/f_\star(t_+)$ by using Eq.\ \eqref{fstarratio}
and also by numerically solving Eqs.\ \eqref{f*dot} $-$
\eqref{eta*dot}; for the numerical calculation, we adopt the setup
given by Eqs.\ \eqref{a0khat}, \eqref{titf}, and \eqref{initialcond}.
The results are shown in Table \ref{table:fstarratio} for several
values of $H$.  We can find an excellent agreement between the
numerical result and the semi-analytic result given in
Eq.\ \eqref{fstarratio}, even though an enormous increase of $f_\star$
occurs while the mode is in the resonance band.

\begin{table}[t]
  \begin{center}
    \begin{tabular}{rrrr}
      \hline\hline
      $h$ & Numerical & Eq.\ \eqref{fstarratio} & Eq.\ \eqref{fstarboltz}
      \\
      \hline
      $2\times 10^{-2}$ & $8.4\times 10^{8}$
      & $3.4\times 10^{8}$ & $5.8\times 10^{16}$
      \\
      $1\times 10^{-2}$ & $1.7\times 10^{17}$
      & $1.1\times 10^{17}$ & $6.4\times 10^{33}$
      \\
      $5\times 10^{-3}$ & $1.0\times 10^{34}$
      & $1.3\times 10^{34}$ & $8.3\times 10^{67}$ 
      \\
      \hline\hline
    \end{tabular}
    \caption{The ratio $f_\star(t_-)/f_\star(t_+)$, based on the
      numerical calculation (Numerical) and Eq.\ \eqref{fstarratio},
      taking $q=10^{-2}$.  We take several values of $q$ and $H$
      (parameterized as $H=hq^2m_\phi$).  For comparison, we also show
      the value of the right-hand side of Eq.\ \eqref{fstarboltz}. }
    \label{table:fstarratio}
  \end{center}
\end{table}

\subsection{Breakdown of Boltzmann equation}

Finally, let us come to our main point. 
We compare the above results with that obtained by using the
Boltzmann equation.  Adopting the collision term given in
Eq.\ \eqref{collisionterm}, the Boltzmann equation with the cosmic
expansion is
\begin{align}
  \dot{f}_k = 
  H k \frac{\partial f_k}{\partial k}
  + 2 \Gamma_{\phi \to \chi \chi}^{(0)} n_\phi \delta (k-p_\chi)
  (1 + 2f_k)
  \left( \frac{p_\chi^2}{2\pi^2} \right)^{-1},
  \label{Boltzmann_cosmo}
\end{align}
or equivalently, 
\begin{align}
  \dot{f}_\star = 
  2 \Gamma_{\phi \to \chi \chi}^{(0)} n_\phi \delta (K-p_\chi)
  (1 + 2f_\star)
  \left( \frac{p_\chi^2}{2\pi^2} \right)^{-1}.
\end{align}
One can easily solve this equation and obtain
\begin{align}
  f_\star (t\rightarrow\infty) =
  \frac{1}{2}
  \left[ \exp \left( \frac{\pi q^2 m_\phi}{4H} \right) - 1 \right].
  \label{fstarboltz}
\end{align}
We can see that, when the effect of the stimulated emission is
effective, the enhancement factor suggested from the Boltzmann
equation is exponentially larger than that from the argument based on
the QFT (see Eq.\ \eqref{fstarratio}); the exponent is doubled in the
result from the Boltzmann equation.  (For comparison, we also show the
right-hand side of \eqref{fstarboltz} in Table
\ref{table:fstarratio}.)  Notice that the discrepancy in the present
case is rather serious than that in the flat spacetime.  This is
because, here, we consider the evolution of the mode with a fixed
comoving momentum which goes through the resonance band; thus the
discrepancy cannot be solved even if we consider a prescription to
average the growth rate in the resonance band.

We may na\"{i}vely set an ansatz for a Boltzmann equation with
including the quantum effect.  We can find that the result based on
the QFT is well described by Eq.\ \eqref{Boltzmann_cosmo} with
replacing $(1 + 2f_k)\rightarrow (1 + f_k)$ in the right-hand side,
i.e.,
\begin{align}
  \dot{f}_k = 
  H k \frac{\partial f_k}{\partial k}
  + 2 \Gamma_{\phi \to \chi \chi}^{(0)} n_\phi \delta (k-p_\chi)
  (1 + f_k)
  \left( \frac{p_\chi^2}{2\pi^2} \right)^{-1}.
  \label{QBE}
\end{align}
The solution of the above equation is given by
\begin{align}
  f_\star (t\rightarrow\infty) =
  \exp \left( \frac{\pi q^2 m_\phi}{8H} \right) - 1,
  \label{fstarboltz2}
\end{align}
which has the same growing behavior as that in the QFT (see
Eq.\ \eqref{fstarratio}).  Interestingly, Eq.\,\eqref{fstarboltz2}
well describes the numerical result 
for any value of $f_\star (t\rightarrow\infty)$.  The theoretical
justification of this equation will be considered in the future
\cite{MoroiYinFuture}.

\section{Summary}
\label{sec:summary}
\setcounter{equation}{0}

In this paper, we have studied particle production from an oscillating
scalar field $\phi$, assuming that the final state particle $\chi$ is
very weakly interacting.  We have paid particular attention to the
consistency of the results from the Boltzmann equation and those from
the QFT calculation.  We have concentrated on the case that the
production of $\chi$ is via the process $\phi\rightarrow\chi\chi$.

First, we have considered particle production in the flat spacetime.
In such a case, we have discussed the evolution of the occupation
number of each mode (i.e., mode with a fixed momentum $k$) separately
in the narrow resonance regime.  We have derived the evolution
equations for the occupation number of each mode based on the QFT.  A
resonance band shows up at $\omega_k$ close to $\frac{1}{2}m_\phi$,
which corresponds to the lowest resonance band in the context of the
parametric resonance.  The modes within the resonance band can be
effectively produced.  For the timescale much longer than
$(qm_\phi)^{-1}$, the occupation numbers of the modes in the resonance
band exponentially grow; the growth rate obtained in our analysis is
consistent with that given by the study of the parametric resonance
using the Mathieu equation.  Then, comparing the occupation number
obtained from the QFT calculation with that from the Boltzmann
equation, we have found that they do not agree well when the
occupation number is larger than $\sim 1$.  On the contrary, when
$f_k\ll 1$, we have found a good agreement of two results.  We have
also argued how our evolution equation based on the QFT could be
related to the ordinary Boltzmann equation.  When the occupation
number is larger than $\sim 1$, some of the approximation and
assumption necessary for such an argument cannot be justified, which,
we expect, causes the disagreement.

Then, we have studied particle production taking into account the
effects of cosmic expansion.  With the cosmic expansion, the physical
momentum redshifts.  The momentum of each mode stays in the resonance
band for a finite amount of time and then exits the resonance band.
The exponential growth of the occupation number occurs only in the
resonance band.  The growth factor has been studied numerically and
analytically, adopting the evolution equations based on the QFT.  The
agreement between numerical and analytical results is excellent.  We
have also analyzed the system by using the conventional Boltzmann
equation and found that the growth rate obtained by solving the
Boltzmann equation is a factor of $2$ larger than that based on the
QFT.  Thus, the occupation number from the Boltzmann equation may
become exponentially larger than that from the QFT, and a na\"ive use
of the conventional Boltzmann equation may result in a significant
overestimation of the number density of $\chi$.\footnote
{The conclusions of \cite{Moroi:2020has}, in which the present authors
  used the conventional Boltzmann equation to discuss the stimulated
  dark matter emission from inflaton decays, do not change.  This is
  because, in \cite{Moroi:2020has}, the abundance of the dark matter
  is fixed by observation and the QFT correction only changes the
  requirements on the model parameters by factors of $O(1)$. }

In this paper, we have considered the production of a bosonic
particle, concentrating on the lowest resonance band of the parametric
resonance.  Consideration of the production of fermionic particles
and the study of the higher resonance bands, as well as the use of the
evolution equations based on the QFT to other phenomena, are left as
future works \cite{MoroiYinFuture}.

\section*{Acknowledgement}
We thank K. Nakayama for carefully reading our manuscript and
providing useful comments and references.  This work is supported by
JSPS KAKENHI grant Nos.\ 16H06490 (TM and WY) and 18K03608 (TM).

\appendix

\section{QFT and Boltzmann Equation}
\label{app:boltzmann}
\setcounter{equation}{0}

In this Appendix, we give a discussion which may indicate a potential
reason of the break down of the Boltzmann equation in the QFT.

We start with
\begin{align}
  \frac{1}{T}\int_{-T/2}^{T/2}{dt e^{\pm i m_\phi t}\dot{g}_k}=
  \mp \frac{1}{T} \int_{-T/2}^{T/2}{  im_\phi e^{\pm i m_\phi t}g_k dt }
  +\frac{1}{T} \left[ e^{\pm i m_\phi t}g_k \right]^{T/2}_{-T/2}.
  \label{integ}
\end{align}
Then, let us assume that $\lim_{T\to \infty}\frac{1}{T}g_k(\pm
T/2)\rightarrow 0$, although it cannot be satisfied as we will see in
the following.  The assumption may imply that the state is
asymptotically an eigenstate of the particle numbers, like in the
perturbation theory of the QFT. Then the second term of the right hand
side of Eq.\ \eqref{integ} vanishes in the limit of
$T\rightarrow\infty$.

Then, we use the following relation obtained from \eqref{gdot}:
\begin{align}
  ( 2k \mp m_\phi  - i \epsilon)
  \int_{-T/2}^{T/2} dt e^{\pm i m_\phi t}g_k =
  - \int_{-T/2}^{T/2} dt
  \left[ \frac{ A \phi (t)}{\omega_k}(1+2f_{{k}}+g_k )
  e^{\pm im_\phi t}
  - i \epsilon e^{\pm i m_\phi t}g_k\right],
\end{align}
or equivalently,
\begin{align}
  \frac{1}{T}\int_{-T/2}^{T/2}{e^{\pm i m_\phi t}g_k dt} =
  -\frac{1}{T}
  \frac{1}{( 2k \mp m_\phi)  - i \epsilon}
  \frac{ A\bar{\phi }}{\omega_k}
  \int_{-T/2}^{T/2} dt (1+2f_{{k}})e^{\pm im_\phi t}\cos m_\phi t
  +O(\epsilon,q^2).
\end{align}
At the end of calculation, we take $\epsilon\rightarrow +0$.  
Using $\lim_{\epsilon \to + 0}\Im (x-i\epsilon)^{-1}=\pi\delta(x)$,
the time average of Eq.\ 
\eqref{fdot} becomes\footnote
{If, on the other hand, $\epsilon<0$, we get an equation which
  decreases $f_k$.}
\begin{align}
  \langle \dot{{f}}_{{k}} \rangle_t =
  \frac{\pi q^2 m_\phi}{16}
  \frac{2p_\chi}{m_\phi}
  \left( 1 +2 \langle{{f}_k\rangle}_t \right)
  \delta( k -m_\phi/2 )+O(q^3).
\end{align}
Thus, if $\frac{1}{T}g_k(T/2)=0$ (with taking $\epsilon>0$), we obtain
the collision term in the Boltzmann equation \eqref{collisionterm}.
However, $\frac{1}{T}g_k(T/2)=0$ does not hold in the limit of
$T\rightarrow\infty$.  To see this, we can use the following quantity:
\begin{align}
  J_k\equiv \left( f_k+\frac{1}{2} \right)^2 - |g_k|^2,
\end{align}
which is time independent, i.e., 
\begin{align}
  \dot{J}_k = 0.
\end{align}
With the initial condition of our choice, $J_k=\frac{1}{4}$.  For $k$
in the resonance band, $\frac{1}{T}g_k(T/2)$ diverges as
$T\rightarrow\infty$ because of the exponential growth of $f_k$.  As a
result, we may conclude that the Boltzmann equation cannot be derived
because the state at $t\rightarrow\infty$ is not an eigenstate of the
particle numbers.


\begin{thebibliography}{99}

\bibitem{Kolb:1990vq}
E.~W.~Kolb and M.~S.~Turner,
``The Early Universe,''
Front. Phys. \textbf{69}, 1-547 (1990)

\bibitem{Mukhanov:2005sc}
  V.~Mukhanov,
  ``Physical Foundations of Cosmology,''
  Cambridge University Press (2005).

\bibitem{Starobinsky:1980te}
A.~A.~Starobinsky,
Adv. Ser. Astrophys. Cosmol. \textbf{3}, 130-133 (1987)

\bibitem{Guth:1980zm}
A.~H.~Guth,
Adv. Ser. Astrophys. Cosmol. \textbf{3}, 139-148 (1987)

\bibitem{Sato:1980yn}
K.~Sato,
Mon. Not. Roy. Astron. Soc. \textbf{195}, 467-479 (1981)
NORDITA-80-29.

\bibitem{Linde:1981mu}
A.~D.~Linde,
Adv. Ser. Astrophys. Cosmol. \textbf{3}, 149-153 (1987)

\bibitem{Albrecht:1982wi}
A.~Albrecht and P.~J.~Steinhardt,
Adv. Ser. Astrophys. Cosmol. \textbf{3}, 158-161 (1987)

\bibitem{Enqvist:2001zp}
K.~Enqvist and M.~S.~Sloth,
Nucl. Phys. B \textbf{626}, 395-409 (2002)
[arXiv:hep-ph/0109214 [hep-ph]].

\bibitem{Lyth:2001nq}
D.~H.~Lyth and D.~Wands,
Phys. Lett. B \textbf{524}, 5-14 (2002)
[arXiv:hep-ph/0110002 [hep-ph]].

\bibitem{Moroi:2001ct}
T.~Moroi and T.~Takahashi,
Phys. Lett. B \textbf{522}, 215-221 (2001)
[erratum: Phys. Lett. B \textbf{539}, 303-303 (2002)]
[arXiv:hep-ph/0110096 [hep-ph]].

\bibitem{Preskill:1982cy}
J.~Preskill, M.~B.~Wise and F.~Wilczek,
Phys. Lett. B \textbf{120}, 127-132 (1983)

\bibitem{Abbott:1982af}
L.~F.~Abbott and P.~Sikivie,
Phys. Lett. B \textbf{120}, 133-136 (1983)

\bibitem{Dine:1982ah}
M.~Dine and W.~Fischler,
Phys. Lett. B \textbf{120}, 137-141 (1983)

\bibitem{Graham:2018jyp}
P.~W.~Graham and A.~Scherlis,
Phys. Rev. D \textbf{98}, no.3, 035017 (2018)
[arXiv:1805.07362 [hep-ph]].

\bibitem{Guth:2018hsa}
F.~Takahashi, W.~Yin and A.~H.~Guth,
Phys. Rev. D \textbf{98}, no.1, 015042 (2018)
[arXiv:1805.08763 [hep-ph]].

\bibitem{Affleck:1984fy}
I.~Affleck and M.~Dine,
Nucl. Phys. B \textbf{249}, 361-380 (1985)

\bibitem{Moroi:2020has}
T.~Moroi and W.~Yin,
[arXiv:2011.09475 [hep-ph]].


\bibitem{Traschen:1990sw}
J.~H.~Traschen and R.~H.~Brandenberger,
Phys. Rev. D \textbf{42}, 2491-2504 (1990)

\bibitem{Kofman:1994rk}
L.~Kofman, A.~D.~Linde and A.~A.~Starobinsky,
Phys. Rev. Lett. \textbf{73}, 3195-3198 (1994)
[arXiv:hep-th/9405187 [hep-th]].

\bibitem{Shtanov:1994ce}
Y.~Shtanov, J.~H.~Traschen and R.~H.~Brandenberger,
Phys. Rev. D \textbf{51}, 5438-5455 (1995)
[arXiv:hep-ph/9407247 [hep-ph]].

\bibitem{Yoshimura:1995gc}
M.~Yoshimura,
Prog. Theor. Phys. \textbf{94}, 873-898 (1995)
[arXiv:hep-th/9506176 [hep-th]].

\bibitem{Kasuya:1996np}
S.~Kasuya and M.~Kawasaki,
Phys. Lett. B \textbf{388}, 686-691 (1996)
doi:10.1016/S0370-2693(96)01216-6
[arXiv:hep-ph/9603317 [hep-ph]].

\bibitem{Kofman:1997yn}
L.~Kofman, A.~D.~Linde and A.~A.~Starobinsky,
Phys. Rev. D \textbf{56}, 3258-3295 (1997)
[arXiv:hep-ph/9704452 [hep-ph]].



\bibitem{Dufaux:2006ee}
J.~F.~Dufaux, G.~N.~Felder, L.~Kofman, M.~Peloso and D.~Podolsky,
JCAP \textbf{07}, 006 (2006)
[arXiv:hep-ph/0602144 [hep-ph]].

\bibitem{Matsumoto:2007rd}
S.~Matsumoto and T.~Moroi,
Phys. Rev. D \textbf{77}, 045014 (2008)
[arXiv:0709.4338 [hep-ph]].

\bibitem{Asaka:2010kv}
T.~Asaka and H.~Nagao,
Prog. Theor. Phys. \textbf{124}, 293-314 (2010)
[arXiv:1004.2125 [hep-ph]].

\bibitem{Mukaida:2013xxa}
K.~Mukaida, K.~Nakayama and M.~Takimoto,
JHEP \textbf{12}, 053 (2013)
[arXiv:1308.4394 [hep-ph]].


\bibitem{Kitajima:2017peg}
N.~Kitajima, T.~Sekiguchi and F.~Takahashi,
Phys. Lett. B \textbf{781}, 684-687 (2018)
[arXiv:1711.06590 [hep-ph]].

\bibitem{Amin:2019qrx}
M.~A.~Amin, J.~Fan, K.~D.~Lozanov and M.~Reece,
Phys. Rev. D \textbf{99}, no.3, 035008 (2019)
[arXiv:1802.00444 [hep-ph]].

\bibitem{Garcia:2018wtq}
M.~A.~G.~Garcia and M.~A.~Amin,
Phys. Rev. D \textbf{98}, no.10, 103504 (2018)
doi:10.1103/PhysRevD.98.103504
[arXiv:1806.01865 [hep-ph]].

\bibitem{Agrawal:2018vin}
P.~Agrawal, N.~Kitajima, M.~Reece, T.~Sekiguchi and F.~Takahashi,
Phys. Lett. B \textbf{801}, 135136 (2020)
[arXiv:1810.07188 [hep-ph]].




\bibitem{Dror:2018pdh}
J.~A.~Dror, K.~Harigaya and V.~Narayan,
Phys. Rev. D \textbf{99}, no.3, 035036 (2019)
[arXiv:1810.07195 [hep-ph]].

\bibitem{Co:2018lka}
R.~T.~Co, A.~Pierce, Z.~Zhang and Y.~Zhao,
Phys. Rev. D \textbf{99}, no.7, 075002 (2019)
[arXiv:1810.07196 [hep-ph]].





\bibitem{Kaneta:2019zgw}
K.~Kaneta, Y.~Mambrini and K.~A.~Olive,
Phys. Rev. D \textbf{99}, no.6, 063508 (2019)
[arXiv:1901.04449 [hep-ph]].

\bibitem{Lozanov:2019jxc}
K.~D.~Lozanov,
[arXiv:1907.04402 [astro-ph.CO]].

\bibitem{Alonso-Alvarez:2019ssa}
G.~Alonso-\'Alvarez, R.~S.~Gupta, J.~Jaeckel and M.~Spannowsky,
JCAP \textbf{03}, 052 (2020)
[arXiv:1911.07885 [hep-ph]].

\bibitem{Mathieu}
  E.~Mathieu, 
  ``M\'{e}moire sur Le Mouvement Vibratoire d'une Membrane de
  forme Elliptique,'' 
  Journal de Math\'{e}matiques Pures et Appliqu\'{e}es, 137-203 (1868). 

\bibitem{Mathieu2}
  M.~William. 
  ``Theory and application of Mathieu functions," 
  Oxford University Press (1951).

\bibitem{MoroiYinFuture}
  T.~Moroi and W.~Yin,
  work in progress.


\end{thebibliography}
\end{document}